\documentclass{kapproc}
\usepackage{amsmath}
\usepackage{epsfig}
\usepackage{mathptm}
\usepackage{procps}

\normallatexbib

\begin{document}

\articletitle[Phase dependent current statistics in
short-arm Andreev interferometer]{Phase dependent current\\
statistics in short-arm Andreev\\ interferometer}

\author{E.~V.~Bezuglyi$^{1,3}$ \\ E.~N.~Bratus'$^{1}$}

\affil{Institute for Low Temperature Physics and Engineering, Kharkov
61103, Ukraine$^{1}$}
\author{V.~S.~Shumeiko$^{2}$}
\affil{Chalmers University of Technology, Gothenburg S-41296,
Sweden$^{2}$}
\author{V.~Vinokur$^{3}$}
\affil{Argonne National Laboratory, Argonne IL 60439, U.S.A.$^{3}$}

\begin{abstract}

We calculate analytically the current statistics for a short
diffusive wire between the normal reservoir and a short
superconductor-normal metal-superconduc\-tor (SNS) junction, at
arbitrary applied voltages and temperatures. The cumulant-generating
function oscillates with the phase difference $\phi$ across the
junction, approaching the normal-state value at $\phi=\pi$. At $T=0$
and at the applied voltage much smaller than the proximity gap
$\Delta_\phi$, the current noise $P_I$ doubles and the third current
cumulant $C_3$ is 4 times larger with respect to their values in the
normal state; at $eV \gg \Delta_\phi$ they acquire large excess
components. At the gap edge, $eV = \Delta_\phi$, the effective
transferred charge defined through $dP_I/dI$ and $dP_I/dV$ approaches
$0e$ and $3e$, respectively, which makes doubtful the interpretation
of these quantities as physical elementary transferred charge. At $T
\neq 0$, $C_3$ shows a non-monotonous voltage dependence with a dip
near $eV = \Delta_\phi$.

\end{abstract}

During last years, statistics of quantum and thermal fluctuations of
the electric current in mesoscopic systems attracted rapidly growing
attention. It was established both experimentally and theoretically
that the fluctuation properties of mesoscopic conductors provide
important information about correlations and statistics of charge
carriers which are not accessible through conductance measurements. A
powerful theoretical approach to the fluctuation problem has been
developed by using the concept of full counting statistics (FCS),
i.e., statistics of number of particles transferred through the
conductor. The concept of FCS, which first appeared in quantum
optics, was extended to normal electron systems in Ref.~\cite{LLL},
and then applied to superconducting structures in Ref.~\cite{Muz}.

The basic problem of the FCS is to calculate a probability
$P_{t_0}(N)$ for $N$ particles to pass a system during an observation
time $t_0$. Equi\-valently, one can find a cumulant generating
function (CGF) $S(\chi)$,
\begin{equation}
\exp[-S(\chi)] = \sum\nolimits_N P_{t_0}(N) \exp(iN\chi),
\label{CGF}
\end{equation}
which determines the irreducible current correlation functions
(cumulants) $C_n$,
\begin{equation} \label{Cn}
C_n = - \left.(\partial / i\partial \chi)^n S(\chi)\right|_{\chi=0}.
\end{equation}
The first two cumulants, $C_1 =\overline{N}\equiv \sum\nolimits_N N
P_{t_0}(N)$ and $C_2 = \overline{(N - \overline{N})^2}$, correspond
to the average current $I = (e/t_0)C_1$ and noise power $P_I =
(2e^2/t_0)C_2$. Intensive studies of the current noise have led to a
number of interesting results concerning statistical correlations in
the current transport (for a review, see Ref.~\cite{BB}), and an
effective charge $q_{\it eff}$ transferred during an elementary
transport event. The third cumulant $C_3 = \overline{(N
-\overline{N})^3}$ has recently attracted a special interest as the
lowest-order correlation function which is not masked by equilibrium
fluctuations \cite{Gut}. First measurements of $C_3(V)$ in the tunnel
junction \cite{ReuletT} have revealed its high sensitivity to an
electromagnetic environment \cite{BKN}.

In normal metal/superconducting (NS) hybrid structures, the basic
mechanism of charge transport at subgap energies, $E<\Delta$, is due
to Andreev reflection of quasiparticles at the NS boundary
\cite{Andreev}, i.e., conversion of electrons incident from the
normal metal to retroreflected holes, accompanied by escape of Cooper
pairs into the superconductor. During an elementary Andreev
reflection event, the effective charge transferred through the NS
interface is twice the electron charge, $q_{\it eff} = 2e$. This
charge doubling strongly affects the current statistics in NS
junctions; in particular, it is known as the reason for the doubling,
compared to normal junctions, of a zero-bias shot noise
\cite{Muz,doubling}. At finite bias, the effective charge becomes
dependent on applied voltage \cite{Naz2,phasenoise}, due to
variations of the size of the proximity region near the NS boundary,
where the quantum coherence holds between the electrons and
retroreflected holes.

\begin{figure}[tb]
\centerline{\epsfxsize=7.5cm\epsffile{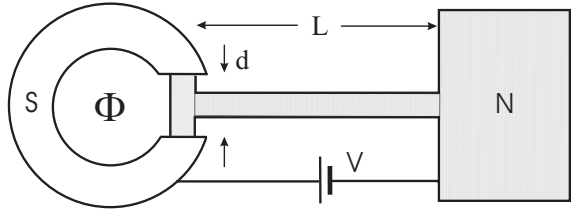}} \vspace{0mm}
\caption{A model of Andreev interferometer: diffusive wire of the
length $L$ connects a normal reservoir (N) and short SNS junction of
the length $d$; magnetic flux $\Phi$ threads a superconducting loop
(S) of the interferometer.}
\label{model}\vspace{-3mm}
\end{figure}

In the Andreev interferometers (see Fig.~\ref{model}), the phase
relations between the electron and hole wavefunctions in the normal
wire can be controlled by magnetic flux enclosed by a superconducting
loop, which results in a periodical dependence of the transport
characteristics of the interferometer on the superconducting phase
difference $\phi$ across the SNS junction. First, the oscillations of
the conductance have been investigated experimentally (see a review
in Ref.~\cite{InterExp}) and theoretically \cite{InterTheor}, and,
recently, the oscillations of the current noise were reported
\cite{phasenoise}.

Motivated by growing interest in higher correlation functions, we
discuss in this paper the full statistics of charge transport in
Andreev interferometers. We will adopt several simplifying
assumptions, which enables us to present an analytical solution for
the CGF and thus to clearly demonstrate essential features of
coherent effects in the current statistics in NS structures. Our
approach is based on the extended Keldysh-Green technique
\cite{NazFCS} (see also \cite{BelzigRev}) in which the CGF is
determined by the equation
\begin{equation} \label{EqS}
(-ie/t_0)\partial S / \partial \chi = I(\chi),\;\;
 I(\chi) = {1\over 8e} \int dE\, {\rm{Tr}}\, \check\tau_K
\check{I},\quad  \check{\tau}_K = \sigma_z\tau_x.
\end{equation}
The Pauli matrices $\sigma$ ($\tau$) operate in the Nambu (Keldysh)
space. The counting current $I(\chi)$ is to be found from the quantum
kinetic equations \cite{LO} for the $4\times 4$ matrix Keldysh-Green
function $\check{G}$ in the mesoscopic normal region of the
interferometer confined between the reservoirs,
\begin{eqnarray} \label{Keldysh}
&\displaystyle \sigma_N \left[\sigma_z E,\check{G}\right]=i
\hbar{\cal D}\partial \check{I}, \quad \check{I} = \sigma_N \check{G}
\partial \check{G}, \quad {\check{G}}^2=\check{1}.
\end{eqnarray}
In this equation, $\cal D$ is the diffusion coefficient,
$\partial$ denotes spatial derivative, and $\sigma_N$ is the
normal conductivity per unit length. The counting field $\chi$ is
introduced via a modified boundary condition involving the gauge
transformation of the lo\-cal-equi\-librium function
$\check{G}_R$, e.g., in the right ($R$) normal reservoir,
\begin{equation} \label{gauge}
\check{G}_R(\chi) = \exp({i\chi\check{\tau}_K/2}) \check{G}_R
\exp({-i\chi\check{\tau}_K/2}).
\end{equation}
A brief overview of this technique in the particular case of
normal structures can be found in the Appendix.

For a multi-terminal structure in Fig.~\ref{model}, the solution of
Eq.~(\ref{Keldysh}) has to be found separately in each arm of the
interferometer, with the matching condition following from the
Kirchhoff's rule for partial counting currents at the node
\cite{NazMulti}. The problem simplifies when the junction length $d$
is much smaller than the length $L$ of the interferometer wire (or,
more precisely, the wire resistance dominates the net interferometer
resistance). In this case, the interferometer wire weakly affects the
spectrum of the junction \cite{3T}, which thus can be considered as
an effective left ($L$) reservoir. Correspondingly, the function
$\check{G}_L$ which imposes the boundary condition to
Eq.~(\ref{Keldysh}) at the junction node, is to be constructed from
the Green's and distribution functions at the middle of a closed
equilibrium SNS junction. Furthermore, if $d$ is much smaller than
the coherence length $\xi_0 = \sqrt{\hbar{\cal D}/\Delta}$, these
Green's functions take the BCS form, with the phase-de\-pen\-dent
proximity gap $\Delta_\phi = \Delta |\cos(\phi/2)|$ \cite{KO}. Within
such model, the problem of current statistics in the Andreev
interferometer reduces to the calculation of the CGF for an NS
junction with the effective order parameter $\Delta_\phi$ in the
superconducting reservoir.

Proceeding to this calculation, we encounter a common difficulty,
violation of the standard triangle form of $\check{G}$ in the Keldysh
space resulting from the gauge transformation in Eq.~(\ref{gauge}).
In such situation, Eq.~(\ref{Keldysh}) cannot be decomposed into the
Usadel equation for the Green's functions and the kinetic equation
for the distribution functions, which makes well developed methods
for solving Keldysh-Green's equations quite unusable. For this
reason, the FCS problem in NS structures generally requires a
numerical analysis of the whole $4\times4$ matrix boundary problem
which has been so far performed only in the limit of small
characteristic energies $\{eV,T\} \ll \Delta$ \cite{Naz2,phasenoise}.

In some particular cases, an analytical solution of this problem can
be obtained by means of the generalized circuit theory
\cite{NazT,NazSIS}. Within this approach, the CGF for a mesoscopic
connector between two reservoirs is expressed in terms of the
distribution $\rho({\mathrm T})$ of the transparencies of the
conduction channels,
\begin{eqnarray}\label {NI}
&\displaystyle S(\chi) ={gt_0\over 4e^2} \int \!dE \int^1_0
\!d{\mathrm T} \rho ({\mathrm T})\,{\rm Tr }\, \ln \check{W}(E,{\mathrm T},\chi), \\
\label{anticomm} &\displaystyle  \check{W}=1 + ({{\mathrm T}/ 4}) (\{
\check{G}_L , \check{G}_R(\chi) \} - 2),
\end{eqnarray}
where $g$ is the connector conductivity. Equation (\ref{NI})
generally applies to the normally conducting structures with
arbitrary $\rho({\mathrm T})$. It was also applied to the
superconducting tunnel junctions \cite{Bruder} and point contacts
\cite{NazSIS,SIS} with a singular transparency distribution
localized at the junction transparency. In general NS structures,
due to dephasing between the electron and hole wavefunctions
described by the commutator term in Eq.~(\ref{Keldysh}),
statistics of conducting modes, in contrast to normal structures
\cite{Mello,NazUn}, does not reduce to statistics of
transparencies but concerns full scattering matrices. However, if
the characteristic energies are much smaller than the Thouless
energy, $\{eV,T\} \ll E_{\it{Th}} = \hbar {\mathcal D}/L^2$, the
dephasing term in Eq.~(\ref{Keldysh}) can be neglected, which
makes it possible to apply the transparency statistics for a
normal wire \cite{Mello} to the superconducting structure. In long
diffusive junctions, $L \gg \xi_0$, where the Thouless energy is
small, $E_{\it{Th}} \ll \Delta$, the quasiparticle spectrum is
structureless at small energies, $E \ll E_{\it{Th}}$, which
results in linear voltage dependence of the CGF and,
correspondingly, of all cumulants at $eV \ll E_{\it{Th}}$
\cite{BelzigRev}. In the opposite limit, $eV \gg E_{\it{Th}}$, the
CGF for a long junction can be found within so-called
``incoherent'' approximation \cite{incoherent}, by neglecting the
contribution of the coherent proximity region. The calculations in
\cite{BelzigRev,incoherent} lead to the conclusion that the FCS
exhibits the reentrance effect: In both limits, $eV \ll
E_{\it{Th}}$ and $eV \gg E_{\it{Th}}$, it is described by the same
expression for $S(\chi)$. An interesting situation occurs in NS
junctions with opaque interfaces dominating the net resistance of
the junction \cite{NINIS}. In this case, the crossover between the
coherent and incoherent transport regimes occurs at very small
voltage of the order of the inverse dwell time of quasiparticles
confined between the interface barriers.

In this paper, we focus on the case of short NS junctions with the
length smaller than $\xi_0$ and, correspondingly, with large Thouless
energy, $E_{\it{Th}} \gg \Delta_\phi$. In such situation, the energy
region of negligibly small dephasing, $E \ll E_{\it{Th}}$, overlaps
with the region $E \gg \Delta_\phi$, in which the NS junction behaves
as the normal system. This enables us to apply Eq.~(\ref{NI}) and the
transparency statistics for diffusive normal conductor at arbitrary
voltages and temperatures, and present analytical solution of the
problem.

The calculation of the integrand in Eq.~(\ref{NI}) is briefly
discussed below. The Keldysh-Green's function $G_R(\chi)$ in the
normal reservoir is traceless in the Keldysh space and therefore it
can be expanded over the Pauli matrices $\tau$ as
\begin{equation} \label{GR}
\check{G}_R(\chi) = {\vec{\tau}} (\vec{g}_1 + \sigma_z
\vec{g}_z),\quad \vec{g}_1 \vec{g}_z = 0,\quad \vec{g}_1^2 +
\vec{g}_z^2 =1, \quad \vec{\tau} =(\tau_x, \tau_y, \tau_z),
\end{equation}
where the vectors $\vec{g}_{1,z}(\chi)$ are expressed through
local-equilibrium distribution functions in the voltage biased
electrode. In the subgap energy region, $E<\Delta_\phi$, the
function $\check{G}_L$ at the junction node is the unity matrix in
the Keldysh space proportional to the Nambu matrix Green's
function $\hat g$ in the superconductor,
\begin{equation} \label{GL}
\check{G}_L = \hat{g} = \sigma_y \exp(\sigma_x \theta_\phi), \quad
\hat{g}^2 =1, \quad \theta_\phi = {\rm arctanh} \,(E/\Delta_\phi).
\end{equation}
Then the calculation of the trace in the Nambu space in
Eq.~(\ref{anticomm}) is reduced to the summation over the
eigenvalues $\sigma = \pm 1$ of the matrix $\hat{g}$,
\begin{align} \label{TrNambu}
{\mathrm {Tr}} \ln \check{W} &= \sum\nolimits_\sigma {\mathrm
{Tr}}_\tau \ln \check{W}_\sigma, \;\; \check{W}_\sigma= a +
\vec{\tau}\vec{b}, \\
 a = 1 &- {\mathrm T}/ 2,\;\; \vec{b} = {({\mathrm T}/
2)}(\sigma\vec{g}_1 -i\vec{g}_z\sinh\theta_\phi).
\end{align}
Noticing that any $2\times 2$ matrix can be presented in exponential
form as
\begin{align} \label{ab}
 \check{W}_\sigma &= \exp(\ln w + \varphi \check{p}),
\\
\label{ab1}   w^2 &= a^2 - \vec{b}^2,\;\; \cosh \varphi = {a / w},
\;\; \check{p} = {\vec{\tau}\vec{b} / w \sinh\varphi},\;\;
{\mathrm {Tr}}\, \check{p}=0,
\end{align}
where $w$ is independent of $\sigma$ due to orthogonality of the
vectors $\vec{g}_1$ and $\vec{g}_z$, one easily obtains ${\mathrm
{Tr}}_\tau \ln \check{W}_\sigma = \ln w^2$ and ${\mathrm {Tr}} \ln
\check{W} = 2\ln w^2$. At $E>\Delta_\phi$, the function $G_L$ is
traceless in the Keldysh space,
\begin{equation} \label{GL>}
\check{G}_L = \hat{g}(\vec{\tau}\vec{g}_L), \;\; \hat{g} =
\sigma_z \exp(\sigma_x \theta_\phi),  \;\; \theta_\phi = {\rm
arctanh} \,(\Delta_\phi/E),
\end{equation}
where the vector $\vec{g}_L$ is constructed from the equilibrium
distribution function at zero potential. In this case, the $4
\times 4$ matrix $\check{W}$ has the form $\check{W}=a +
\vec{\sigma}\vec{b}$, where $a$ and $\vec{b}^2$ are scalars,
\begin{align} \label{W>}
 a &= 1-({\mathrm T}/ 2)(1-\vec{g}_L\vec{g_z}\cosh
\theta_\phi),
\\
 \vec{b}^2 &= ({\mathrm T}/ 2)^2
[(\vec{g}_L\vec{g_1})^2 - (\vec{g}_L \times \vec{g_z})^2 \sinh^2
\theta_\phi],
\end{align}
therefore it can also be transformed to the matrix exponent
similar to Eqs.~(\ref{ab}) and (\ref{ab1}), with the traceless
matrix $\check{p}= \vec{\sigma}\vec{b}/ w \sinh\varphi$. Following
this line, we obtain ${\mathrm {Tr}} \ln \check{W} = 2\ln w^2$,
and then, performing integration over $\mathrm T$ in
Eq.~(\ref{NI}), we arrive at the final expression for the CGF,
\begin{eqnarray}
&\displaystyle S(\chi) = {gt_0 \over 4e^2} \int_0^{\infty} dE\,
S(E,\chi),\label{SS} \quad S(E,\chi) = \left\{
\begin{array}{ccc} 2\theta^2 , &E < \Delta_\phi,
\\
\!\!\theta_+^2  + \theta_-^2 , &E>\Delta_\phi,
\end{array} \right. \label{SS1}
\end{eqnarray}
where the quantities $\theta$ and $\theta_\pm$ are given by explicit
relations,
%
\begin{align} \label{Z}
Z(0)\cosh^2 \theta &= Z(2\chi)\cosh^2 \theta_\phi  ,
\\
Z(0)\cosh \theta_\pm &= \left[Z(\chi)+ \cos
\chi-1\right]\cosh\theta_\phi \pm \tanh{\epsilon\over 2} [\sinh{p}
-
\\
\nonumber &\sinh\left({p}-i\chi\right)-i\sin \chi ] \sqrt{1-{
\cosh\epsilon+1 \over \cosh p-1}\sinh^2\theta_\phi},
\\
\theta_\phi &= {\rm arctanh}\! \left[\left({\Delta_\phi /
E}\right)^{{\rm sgn}\,(E-\Delta_\phi)}\right], \\
Z(\chi) &= \cosh(\epsilon)+\cosh (p-i\chi), \;\; \epsilon = E/T,
\;\; p = eV/T.
\label{thetapm}
\end{align}
%

By using Eqs.~(\ref{Cn}) and (\ref{SS1})-(\ref{thetapm}), one can
obtain analytical expressions for all cumulants. At zero temperature,
the calculation essentially simplifies. Indeed, at $T \to 0$ and
$E>eV$, the dominating terms in Eqs.~(\ref{Z})-(\ref{thetapm}) are
proportional to $\exp(\epsilon)$, and therefore $\theta$ and
$\theta_\pm$ are equal to $\theta_\phi$. This implies that the CGF is
independent of the counting field at these energies, and all
cumulants turn to zero. At $E<eV$, the terms with $\exp(p-in\chi)$
dominate, and we obtain
\begin{equation} \label{T0}
\cosh\theta = e^{-i\chi}\cosh \theta_\phi, \;\; \cosh\theta_\pm =
e^{-i\chi} \cosh\theta_\phi \pm (e^{-i\chi}-1).
\end{equation}
At subgap voltage, $eV < \Delta_\phi$, when the charge transport at
$T=0$ is only due to the Andreev reflection, the current $I$, the
shot noise power $P_I$, and the third cumulant $C_3$ read
\begin{align} \label{Corr<}
 I &= I_\Delta q(z), \;\; q(z)=\!\int_0^z\! {dx \over x}
\mathop{\mathrm {arctanh}}x, \;\; P_I = 2e\!\left[I - I_\Delta
f(z^{-1})\right],
\\ \label{Corr<1}
 C_3 &= \overline{N} - {\overline{N}_\Delta \over 2z^2 }
\left[ (5z^2 -3 ) f(z^{-1}) + z \right]\!,
 \;\; I_\Delta = {g\Delta_\phi\over e}, \;\; \overline{N}_\Delta
 = {I_\Delta t_0 \over e},
\\
 f(z) &= (1/2) [z - (z^2 -1) \mathop{\mathrm {arctanh}}
z^{-1}], \quad  z = {eV / \Delta_\phi}.
\end{align}
At small voltage, $eV \ll \Delta_\phi$, the shot noise doubles, $P =
(4/3)eI$, and $C_3 = 4\overline{N}/15$  is four times larger compared
to the normal case \cite{doubling,Muz,1/3,LLL}. When the voltage
increases and exceeds the gap edge, $eV > \Delta_\phi$, the normal
electron processes at the energies $E > \Delta_\phi$ begin to
contribute to the charge transport, providing the normal-state
voltage dependencies of the cumulants at $eV \gg \Delta_\phi$. At
large voltage, the Andreev reflected particles produce
voltage-in\-de\-pen\-dent excess components of the cumulants (in
ballistic NS junctions, existence of the excess noise was first
predicted in Ref.~\cite{Khlus}),
\begin{eqnarray} \label{Corr>} &\displaystyle \!\!\!\!\!\!I =
I_N  - I_\Delta f(z) + I^{\rm{ex}}, \;\; P_I = 2e I_\Delta (z^2 - 1)
f(z) + P_I^{\rm{ex}}, \;\; I_N = gV,
\\
&\displaystyle \!\!\!\!\!\!C_3 = {\overline{N}_\Delta \over 2} (z+1)
\{(z-1)[ 8z/ 3 -  (8z^2 -3)f(z)] -1/3 \} + C_3^{\rm{ex}},
\\
&\displaystyle \!\!\!\!\!\ I^{\rm{ex}} = {I_\Delta \over 2 }
\left({\pi^2 \over 4} -1 \right), \;\; P_I^{\rm{ex}} =
2eI^{\rm{ex}},\;\;
C_3^{\rm{ex}} = {\overline{N}_\Delta \over 2 } \left({\pi^2 \over 4}
- {4\over 3} \right).
\end{eqnarray}
\begin{figure}[tb]
\centerline{\epsfxsize=9cm\epsffile{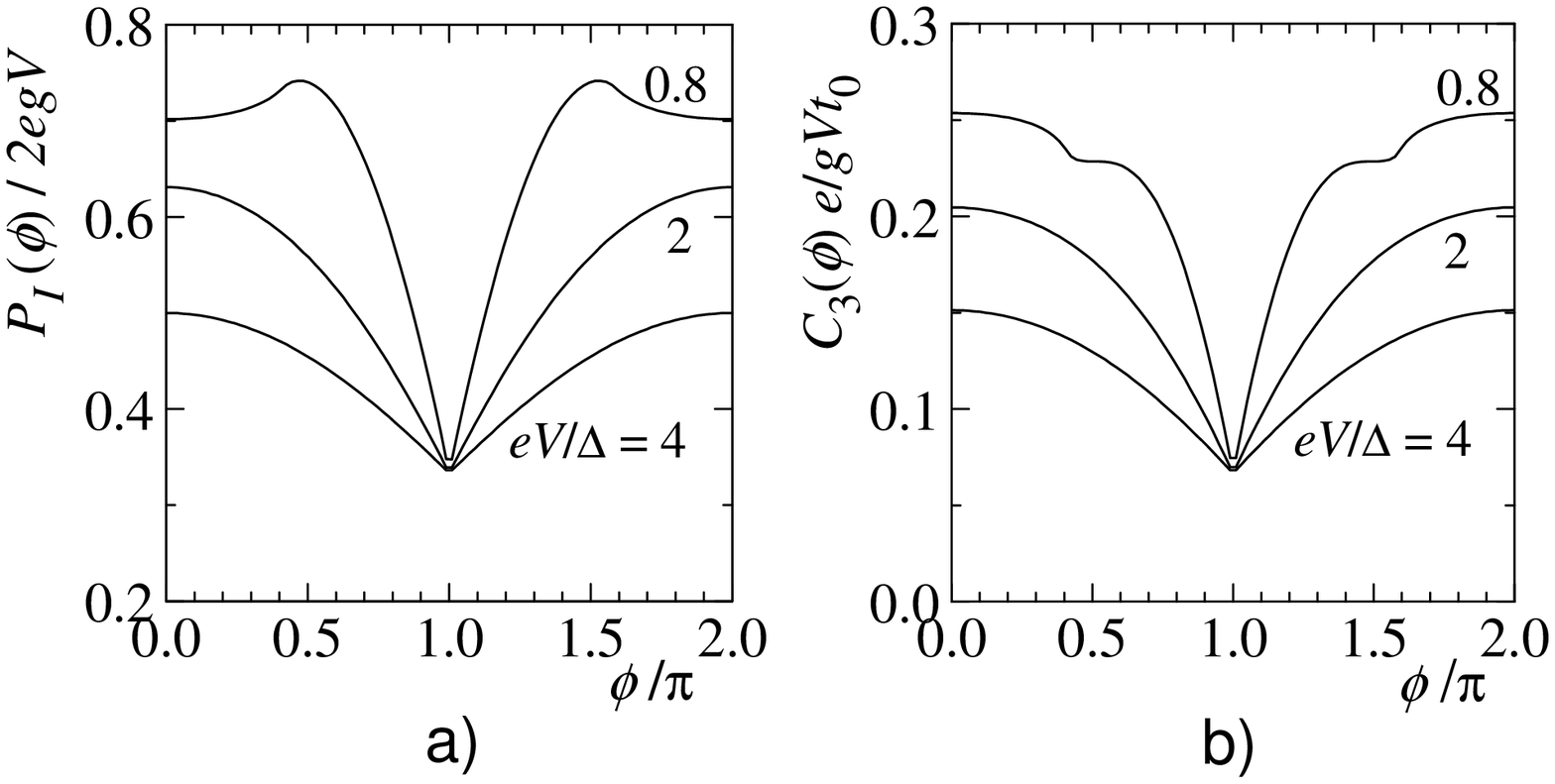}}\vspace{-2mm}
\vspace{2mm} \centerline{\epsfxsize=9cm\epsffile{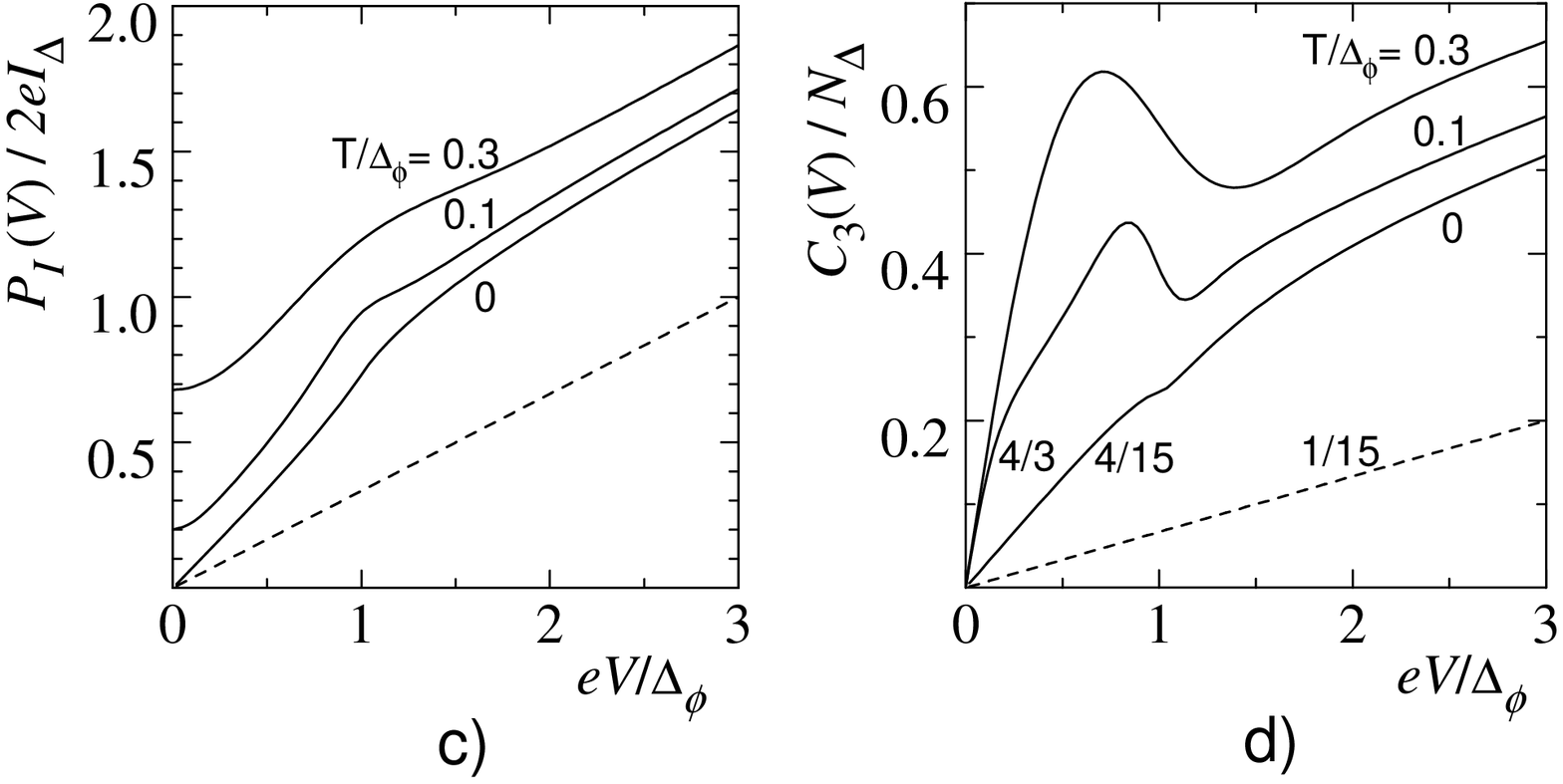}}
\vspace{-3mm}
\caption{Shot noise power and third cumulant vs supercon\-duc\-ting
phase (a,b) at different vol\-tages and $T=0$, and vs voltage at
different temperatures (c,d). Dashed lines denote voltage
dependencies in the normal state at $T=0$. In the panel (d),
zero-bias slopes of the normalized $C_3(V)$ are indicated.}
\label{PC(V)}\vspace{-4mm}
\end{figure}

At nonzero temperatures, $T \neq 0$, we calculate the cumulant
spectral densities $I(E)$, $P(E)$ and $C(E)$ defined as
\begin{equation} \label{Spectral}
I= I_\Delta \!\!\int_0^\infty \!\!\!dE \, I(E), \; P_I= 2eI_\Delta
\!\! \int_0^\infty \!\!\!dE \, P(E), \; C_3= \overline{N}_\Delta \!\!
\int_0^\infty \!\!\!dE \, C(E).
\end{equation}
Here $I(E)\! =\!\! f_1\! \sinh p /Z(0)$, and the functions $P(E)$ and
$C(E)$ at $E\!<\!\Delta_\phi$ read
\begin{align} \label{Densities}
P(E) &= {2 \over Z^2(0)} \left[ 2Q f_1 + (1-f_2) \sinh^2 p
\right],\;\; Q = 1 + \cosh\epsilon \cosh p,
\\
C(E) &= {\sinh p \over Z^3(0)} \left[ 4f_1 \sinh^2 \epsilon +
(2f_2\! +\! 3f_3) \sinh^2 p\! +\! 2Q \Big(3(1\! -\! f_2)\!-\! 2
f_1\Big) \right]\!\!,
\end{align}
whereas at $E > \Delta_\phi$ they are given by equations,
\begin{align} \label{C>}
 P(E) &= {2 \over Z^2(0)} \left[ Q\left(1+2f_1 - 2f_2
{\cosh p - 1\over \cosh \epsilon +1} \right) +  \sinh^2 p -Z(0)
\right],
\\
C(E) &= {\sinh p \over Z^3(0) (1+\cosh \epsilon)}
\Big\{4f_1(1+\cosh \epsilon)(Q+\sinh^2 \epsilon) +
\\ \nonumber
&3\Big[ Z(0)(1-2f_3)+Q \Big(4(1-f_2+f_3 \cosh\epsilon ) + 3
\cosh\epsilon - 2f_3\Big) +
\\ \nonumber
\sinh^2 &\epsilon \Big(2f_3 - \cosh\epsilon + \left.\left. (3-5
\cosh p )f_2 \Big)\right] + f_2(5 \cosh \epsilon -1) \sinh^2 p
\right\},
\\
f_1 = & \theta_\phi \coth \theta_\phi, \;\; f_2 = {(f_1\! -1) /
\sinh^2 \theta_\phi}, \;\; f_3 = {(f_2\! -1/3) / \sinh^2
\theta_\phi}. \label{f}
\end{align}
In Eqs.~(\ref{Densities})-(\ref{f}), the functions $f_i$ describe
energy variation of quasiparticle spectrum which is most essential in
the vicinity of the gap edge $\Delta_\phi$.

As shown in Fig.~\ref{PC(V)},(a,b), the cumulants oscillate with the
phase and exhibit deep minima at $\phi \,{\rm{mod}}\, 2\pi = \pi$,
when the gap closes and the cumulants approach their normal values.
When the proximity gap $\Delta_\phi$ approaches $eV$, $P_I(\phi)$
exhibits a peak, while $C_3(\phi)$ shows a step-like structure. The
voltage dependence of the cumulants for different temperatures is
plotted in Fig.~\ref{PC(V)},(c,d) in specifically normalized
variables, which provides universality of the curves for any $\phi$.
As the temperature increases, the current noise approaches finite
value at $eV=0$ due to thermal fluctuations, and exhibits quadratic
dependence on the applied voltage at $eV \ll T$. Within the
intermediate voltage region, $T < eV < \Delta_\phi$, $P_I(V)$ becomes
linear with doubled slope produced by the Andreev reflected
particles, and at $eV > \Delta_\phi$, the slope turns to its
normal-metal value. A considerable excess noise at large voltages is
contributed by both the thermal fluctuations and Andreev reflection.
A more interesting behavior is discovered for the third cumulant. As
the temperature departs from zero, the zero-bias slope of the
normalized $C_3(V)$ increases by factor of 5 compared to the zero
temperature (which is similar to the normal structure \cite{Gut}),
and approaches the value $4/3$. Then, at $T < eV < \Delta_\phi$, the
slope of the normalized $C_3(V)$ returns to the subgap value $4/15$
for $T=0$. At $eV \sim \Delta_\phi$, the curve $C_3(V)$ shows ${\cal
N}$-like feature, and finally, at $eV > \Delta_\phi$, it approaches a
straight line with the (normal-state) slope $1/15$. This implies that
$C_3$ acquires anomalously large thermal component at voltage $eV
\sim \Delta_\phi$, which, however, rapidly decreases at $eV >
\Delta_\phi$ and/or $T>\Delta_\phi$ towards the normal metal level.

\begin{figure}[tb]
\centerline{\epsfxsize=8.5cm\epsffile{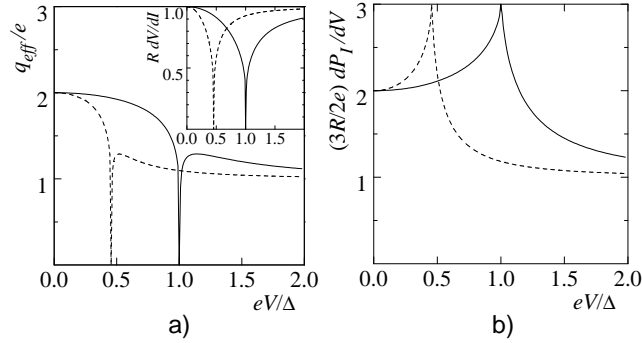}} \vspace{-4mm}
\caption{Effective transferred charge (a), differential noise (b),
and differential resistance (inset) vs voltage at $\phi\!=\!0$ (solid
lines) and $\phi\!=\!0.7\pi$ (dashed lines), $T\!=\!0$. }
\label{diff}\vspace{-4mm}
\end{figure}

The voltage dependence of the effective transferred charge defined
as $q_{\it eff} = (3/2) dP_I/dI$ \cite{phasenoise} and the
normalized differential noise $(3R/2)$ $dP_I/dV$ is shown in
Fig.~\ref{diff} for $T=0$. At zero voltage, both quantities
approach $2e$, whereas at $eV \gg \Delta_\phi$ they turn to $e$.
However, the behavior of these quantities becomes very different
when the voltage approaches the proximity gap edge $\Delta_\phi$:
the normalized differential noise increases to $3e$, whereas the
effective charge $q_{\it eff} \sim (dP_I/dV)(dV/dI)$ rapidly turns
to zero, along with the differential re\-sis\-tan\-ce $dV/dI$ (see
inset in Fig.~\ref{diff}). We note that the effect of zero
differential re\-sis\-tance giving $q_{\it eff} = 0$ results from
the resonant transparency enhancement at the proximity gap edge,
due to the singularity in the density of states. Such an
``exotic'' behavior of these quantities at the gap edge leads us
to the conclusion that none of them can be unambiguously
associated with the physical elementary transferred charge, in
contrast to what is commonly suggested. Similar effects have been
predicted in Ref.~\cite{NINIS} for an NS structure with opaque
interfaces; however, in such case, a considerable enhancement of
$dP_I/dV$ and suppression of $dP_I/dI$ occur at small applied
voltage of the order of the inverse dwell time of quasiparticles.

It is instructive to compare our analytical results for short-arm
interferometers with that obtained numerically for long NS
junctions with a small minigap $E_g \sim E_{\it{Th}}  \ll \Delta$.
The results are qualitatively similar: in long junctions, $q_{\it
eff}$ is equal to $2e$ at $eV \ll E_g$ and has a minimum at $eV
\approx E_g(\phi)$, which moves towards small voltage when at
$\phi \to \pi$ \cite{phasenoise}; the differential noise is also
non-monotonous and approaches maximum at $eV \approx 5E_{\it{Th}}$
\cite{Naz2}. After this comparison we see that the proximity gap
$\Delta_\phi$ in short junctions plays the role of the minigap
$E_g$ in long junctions and determines the feature in the
effective charge, though this feature at $eV \sim E_g$ in long
junctions is much less pronounced. However, as noted above, a
qualitative difference of long junctions is the existence of an
intermediate incoherent voltage region $E_g \ll eV \ll \Delta$,
where both the effective charge and the normalized differential
noise have the value $2e$, and their crossover to $e$ occurs only
at $eV \geq \Delta$ \cite{NagBut}.

\appendix{}

For reference purposes, in this Appendix we outline the procedure
and summarize the results of calculation of the CGF for a
diffusive connector between normal reservoirs, by using the
extended Keldysh-Green's technique. For generality, we consider a
diffusive wire interrupted by tunnel barriers, which enables us to
present several original results and to examine various limiting
situations.

In normal systems, the matrices $\check{G}$ and $\check{I}$ are
traceless in the Keldysh space and therefore they can be expressed
through 3-vectors with the components diagonal in the Nambu space,
$\check{G} = \vec{g} \vec\tau$, $\check{I} = \vec{I} \vec\tau$,
where $\vec\tau$ is the vector of the matrices $\tau$, and
$\vec{g}^2=1$. Since the left-hand side of Eq.~\eqref{Keldysh}
turns to zero in normal systems, the formal solution of
Eq.~\eqref{Keldysh} for the matrix current density $\check{I}_N$
in each segment of the wire can be easily obtained,
\begin{equation} \label{matrixIN}
\check{I}_N = g_N \ln \check{G}_1 \check{G}_2 = g_N \ln [\vec{g}_1
\vec{g}_2 + i\vec{\tau} (\vec{g}_1 \times \vec{g}_2)] =
\vec{\tau}\vec{I}_N, \quad \vec{I}_N = ig_N \vec{p}\phi_N,
\end{equation}
where $g_N$ is the conductivity of the wire segment,
$\check{G}_{1,2}$ are the Green's functions at the left and right
segment edges, respectively, $\phi_{N} = \arccos \vec{g}_1
\vec{g}_2$ is the angle between the (complex) unit vectors
$\vec{g}_1$ and $\vec{g}_2$, and $\vec{p} = (\vec{g}_L \times
\vec{g}_R)/ \sin\phi_{N}$ is the unit vector perpendicular to
$\vec{g}_1$ and $\vec{g}_2$.

The matrix current $\check{I}_B$ through the tunnel barrier can be
expressed through the Green's functions $\check{G}_{-}$ and
$\check{G}_{+}$ at the left and right sides of the barrier by
using the boundary condition \cite{KL},
\begin{equation} \label{matrixIB}
\check{I}_{B-} = \check{I}_{B+} = {g_B\over 2} [\check{G}_-,
\check{G}_+] = \vec{\tau}\vec{I}_B, \quad \vec{I}_B = ig_B \vec{p}
\sin\phi_B, \quad \phi_{B} = \arccos \vec{g}_- \vec{g}_+,
\end{equation}
where $\vec{p} = (\vec{g}_- \times \vec{g}_+) / \sin\phi_{B}$ and
$g_B$ is the barrier conductivity.

The conservation of the matrix current along the connector,
$\check{I}=\,\mathrm{const}$, following from Eq.~\eqref{Keldysh}
and the boundary condition in Eq.~\eqref{matrixIB}, results in
conservation of the vector current, $\vec{I} = \vec{I}_{N} =
\vec{I}_{B} = \,\text{const}$. This implies that for all elements
of the connector, the unit vectors $\vec{p}$ coincide, therefore
the Green's vectors $\vec{g}$ lye within the same plane and the
vector $\vec{p}$ can be constructed from known Green's vectors
$\vec{g}_{L}$ and $\vec{g}_{R}$ in the reservoirs, $\vec{p}=
(\vec{g}_L \times \vec{g}_R)/\sin \phi$, where $\phi$ is the angle
between $\vec{g}_{L}$ and $\vec{g}_{R}$. From the current
conservation, we also conclude that all barriers and wires are
characterized by a single variable $\eta$,
\begin{equation} \label{Eq1}
g_{B}\sin\phi_{B} = g_{N}\phi_{N} = g\eta = \,\text{const},
\end{equation}
where the normalization constant $g$ is chosen to be equal to the
conductance of the whole connector. Thus, the vector current is
given by equation,
\begin{equation} \label{Idensity}
\vec{I} = {ig\eta \over \sin \phi} (\vec{g}_L \times \vec{g}_R).
\end{equation}
The planar rotation of the Green's vector along the connector
results in the additivity of the angles between all consecutive
vectors $\vec{g}$, therefore the sum of these angles is equal to
$\phi$,
\begin{align} \label{Eq2}
\sum_{\it wires} \phi_N &+ \sum_{\it barriers} \phi_B = \phi =
\arccos(\vec{g}_L\vec{g}_R),
\\
\vec{g}_L\vec{g}_R &= 1 +
P_{-+}\Big(e^{i\chi}-1\Big)+P_{+-}\Big(e^{-i\chi}-1\Big),
\\
P_{\sigma\sigma^\prime} &= n_\sigma(1-n_{\sigma^\prime}), \quad
n_- = n_F(E), \quad n_+ = n_F(E+eV),
\end{align}
which leads to the following equation for the parameter
$\eta(\phi)$,
\begin{equation} \label{Eqeta}
\gamma_N \eta + \sum\nolimits_k \arcsin (\gamma_{k} \eta) = \phi,
\quad \gamma_{N} = {R_N / R}, \quad \gamma_{k} = {R_k / R}, \quad
\gamma_N + \sum\nolimits_k \gamma_{k}  = 1,
\end{equation}
where $R_N$ is the net resistance of all wires, $R_{k}$ is the
resistance of the $k$-th barrier, and $R=g^{-1}$.

By using the definitions in Eq.~\eqref{EqS}, we obtain the
counting electric current $I(\chi)$ and the CGF,
\begin{align} \label{Ichi1}
I(\chi) &= {1\over 2e}\int_0^\infty dE\, \mathop{\mathrm{Tr}}
\sigma_z \vec{I}_x = {ig\over 2e}\int_0^\infty dE\,
\mathop{\mathrm{Tr}} {\sigma_z \eta \over \sin \phi} (\vec{g}_L
\times \vec{g}_R)_x,
\\
\label{S1} S(\chi) &= {gt_0 \over {4e^2}}\int dE \,
\mathop{\mathrm{Tr}} \Big[ {r_N \eta^2 / 2} + \sum\nolimits_k
\Big(1-\sqrt{1-r_k^2 \eta^2}\Big) / r_k \Big],
\end{align}
We note that the statistics is insensitive to the position of the
barriers, and depends only on the barrier resistances and the net
resistance of the diffusive part of the connector. In the absence
of barriers, $r_k\to 0$, the CGF has the form,
\begin{align} \label{Sdiff}
S(\chi) &= {gt_0 \over 4e^2} \int dE \,\phi^2 = {gt_0 \over 4e^2}
\int dE \,  \arccos^2\Big[1 +
P_{-+}\Big(e^{i\chi}-1\Big)+P_{+-}\Big(e^{-i\chi}-1\Big)\Big].
\end{align}
At zero temperature, the integration over energy in Eq.~(\ref{S1})
can be explicitly performed,
\begin{equation} \label{S2}
S(\chi) = {\overline{N} \over 2} \Big[  r_N \eta^2/2 +
\sum\nolimits_k \Big(1-\sqrt{1-r_k^2 \eta^2}\Big) / r_k \Big],
\end{equation}
where $\overline{N} = gVt_0/e$. From Eq.~\eqref{S2} we find the
Fano factor $F$ in the shot noise power $P_I = eFI$,
\begin{equation} \label{F0}
F = (2/ 3) \left( 1 + 2 B_3 \right), \qquad B_n = \sum\nolimits_k
r_k^n,
\end{equation}
which varies between the Poissonian value $F=2$ for the tunnel
connector and $1/3$-suppressed value, $F = 2/3$, in the absence of
barriers. The third cumulant $C_3$ varies between $\overline{N}$
for Poissonian statistics in the single barrier case and
$\overline{N}/15$ for a diffusive conductor,
\begin{equation} \label{C30}
C_3(V,0) = \left(\overline{N} / 15 \right) \left[1+ 10 B_3
\left(1+ 4B_3\right) -36 B_5 \right].
\end{equation}

It is interesting to note that Eq.~\eqref{Eqeta} can be easily
transformed into equation for the transparency distribution
$\rho({\mathrm{T}})$, by making use of the relation of the
generalized circuit theory between the counting current $I(\chi)$
and the matrix current $\check{I}$ following from Eqs.~\eqref{NI}
and \eqref{EqS},
\begin{equation} \label{matrixI}
I(\chi) = {1\over 4e} \int_0^\infty dE \, \mathop{\mathrm{Tr}}
\tau_x \sigma_z \check{I}, \quad \check{I} = {g \over 2} \int_0^1
d {\mathrm{T}} \rho({\mathrm{T}}){\mathrm{T}} [\check{G}_L,
\check{G}_R(\chi)] \check{W}^{-1}.
\end{equation}
Rewriting this equation in the vector representation, comparing it
with Eq.~\eqref{Ichi1}, and introducing the variable $z =
(1/2)(1-\vec{g_L}\vec{g_R})$, we obtain the integral equation for
$\rho({\mathrm{T}})$,
\begin{equation} \label{Eqp}
\int_0^1 { {\mathrm{T}}d{\mathrm{T}} \rho({\mathrm{T}}) \over
1-z{\mathrm{T}}} = {\eta \over 2\sqrt{z(1-z)}},
\end{equation}
where $\eta$ obeys Eq.~\eqref{Eqeta} with the function $\phi
=2\arcsin\sqrt{z}$ in the right-hand side (rhs). The solution of
Eq.~\eqref{Eqp} has the form $\rho({\mathrm{T}}) =
{\mathop{\mathrm{Re}}\eta / 2\pi {\mathrm{T}} \sqrt{
1-{\mathrm{T}} }}$,
%
%
where $\eta(\mathrm{T})$ is the solution of Eq.~\eqref{Eqeta} with
the function $\pi +
2i\mathop{\mathrm{arccosh}}(1/\sqrt{\mathrm{T}})$ in the rhs
\cite{NazUn}.

In some limiting cases, one can obtain an analytical solution of
Eq.\ (\ref{Eqeta}). In particular, if the number $M$ of the
barriers is large, $M \gg 1$, then the resistance of each barrier
is small compared to the net resistance, $R_k \ll R$. In this
case, the approximate solution of Eq.\ (\ref{Eqeta}) is $\eta =
\phi$, and the CGF coincides with that for diffusive wire,
$S(\chi) = \overline{N} \arccos^2 e^{i\chi/2}$. In the tunnel
limit, when the resistance of each barrier much exceeds the net
resistance of diffusive segments, $R_k \gg R_N$, the first term in
Eq.\ (\ref{Eqeta}) can be neglected. Then an analytical expression
for the parameter $\eta$ and the CGF at arbitrary $M$ can be
obtained in the case of equivalent barriers, $r_k = 1/M$,
\begin{equation} \label{M1}
\eta = M\sin {\phi \over M}, \quad S(\chi) = \overline{N} M^2
\sin^2 {\arccos e^{i\chi/2} \over M},
\end{equation}
when the Fano factor is given by $F = (2/ 3) \left(1+ 2/
M^2\right)$. In the limit of large number of the barriers, $M \gg
1$, we return to the ``diffusive'' statistics, while for
single-barrier structure, $M=1$, we obtain pure Poissonian
statistics, $S(\chi)= \overline{N} (e^{i\chi}-1)$.

At arbitrary temperature, the cumulants can be found analytically
by asymptotic expansion in Eqs.~(\ref{Eqeta}), (\ref{S1}) over
small $\eta$ and $\chi$. In particular, the noise power,
\begin{equation} \label{PT}
P_I(V,T) = {4T \over 3R}\left[\left(1+2B_3\right) {p\over 2} \coth
{p\over 2} +2(1-B_3) \right],
\end{equation}
exhibits crossover between the shot noise at $T \ll eV$ and the
Johnson thermal noise $P_T=4T/R$ at large temperature, $T \gg eV$.
The voltage dependence of the third cumulant,
\begin{equation} \label{C3T}
C_3(V,T) = C_3(V,0) + {2\over 5} \overline{N} (1-10B_3^2 +9B_5)
{\sinh p -p \over p \sinh^2 (p/2)} ,
\end{equation}
is linear in both limits and approaches $(\overline{N}/3)(1+2B_3)$
at high temperatures. In the limit of tunnel connector, $B_n = 1$,
the second term in Eq.~(\ref{C3T}) vanishes, and $C_3$ becomes
temperature independent. In the absence of barriers, $B_n = 0$,
Eq.~(\ref{C3T}) reproduces the result of a modified kinetic theory
of fluctuations for a diffusive wire\cite{Gut}.

In order to access FCS in multi-terminal structures, which consist
of a set of connectors attached between several normal electrodes
and a diffusive island (node) with negligibly small resistance,
separate counting fields $\chi_\alpha$ and parameters
$\eta_\alpha$ are to be introduced in each arm\cite{NazMulti},
\begin{equation} \label{Ialpha}
{\vec{I}}_\alpha =  i\xi_\alpha ({\vec{g}}_\alpha \times
{\vec{g}}_c),\qquad \xi_\alpha = {g}_\alpha \eta_\alpha
/\sin\phi_\alpha.
\end{equation}
The quantities $\eta_\alpha$ obey the equations similar to Eq.
(\ref{Eqeta}), with the angles $\phi_\alpha =
\arccos({\vec{g}}_\alpha {\vec{g}}_c)$ in the rhs, where the
Green's vector ${\vec{g}}_c$ at the node can be found from the
current conservation law, $\sum_\alpha \vec{I}_\alpha = 0$,
\begin{equation} \label{gc}
{\vec{g}}_c = {\vec G}/\sqrt{{\vec G}^2}, \quad {\vec G} =
\sum\nolimits_\alpha \xi_\alpha {\vec{g}}_\alpha.
\end{equation}

According to Eq.~(\ref{gc}), the vector ${\vec{g}}_c$ depends on
all counting fields $\chi_\alpha$, which reflects
cross-cor\-re\-la\-tions between the currents in different
connectors. For the system of tunnel con\-nec\-tors, where the
quantities $\xi_\alpha$ are equal to the conductances $g_\alpha$
and therefore become independent of $\chi$, the CGF at zero
temperature can be explicitly evaluated \cite{Bruder},
\begin{equation} \label{Stunn}
S\{\chi\} = \frac{Vt_0}{2e}G \sqrt{1 + 4 \sum\nolimits_\alpha
\overline{g}_V\overline{g}_\alpha (e^{i\chi_\alpha}- 1)}, \quad
\overline{g}_\alpha = g_\alpha /G, \quad G = \sum\nolimits_\beta
g_\beta,
\end{equation}
where the index $V$ denotes the voltage biased electrode.

For arbitrary connectors, the cumulants can be found from
asymptotic solutions of the equations for $\eta_\alpha$ and
${\vec{g}}_c$ at small $\chi_\alpha$. For instance, the partial
current through $\alpha$-th connector is $I_\alpha = Vg_\alpha
\overline{g}_V$, and the Fano factors defined as $F_{\alpha\beta}
= (2ei/I_\alpha)(\partial I_\alpha\{\chi\} /
\partial \chi_\beta)_{\chi = 0}$ read
\begin{equation} \label{multiF}
F_{\alpha\beta} = \Big(2-{4 \over 3}\overline{g}_V
\Big)\delta_{\alpha\beta}- {4\over 3} \overline{g}_\beta
\Big[1+\overline{g}_V (B_{3\alpha} + B_{3\beta}) -B_{3V}(1-
\overline{g}_V)^2\ - \overline{g}_V \sum_{\gamma \neq V}
\overline{g}_\gamma B_{3\gamma} \Big]\!.
\end{equation}

The diagonal elements $F_{\alpha\alpha}$ of the matrix
$F_{\alpha\beta}$ have the meaning of the Fano factors for the
shot noise in $\alpha$-th connector and may vary between $2/3$ and
$2$, approaching Poissonian value $F_{\alpha\alpha} = 2$ for a
large number of the terminals, when the normalized conductances
$\overline{g}_\alpha$ become small. The cross-correlators
$F_{\alpha\beta}$ ($\alpha \neq \beta$) between the currents in
different terminals are negative\cite{NagBut}. In a particular
case of diffusive connectors ($B_\alpha = 0$), Eq.~(\ref{multiF})
reproduces the result of Ref.~\cite{Sukh} for a so-called
star-shaped geometry.

\end{document}